\documentclass[a4paper,11pt]{article}
\usepackage{pos}
\usepackage{float,physics}
\usepackage{hyperref}
\usepackage{multicol}

\usepackage{tikz}
\usepackage{circuitikz}
\usepackage{graphicx}
\usepackage{animate}
\usepackage{braket}
\usepackage[compat=1.1.0]{tikz-feynman}
\usetikzlibrary{matrix}

\newcommand{\Eqref}[1]{(\ref{#1})}
\newcommand{\Figref}[1]{Fig.~\ref{#1}}

\newcommand{\Ns}{N_s}
\newcommand{\Nt}{N_t}

\title{Fractional instantons and Confinement: first results on a $T_2\times R^2$ roadmap}
\ShortTitle{Fractional instantons and confinement on $T_2\times R^2$}

\author[a]{G.~Bergner}
\author*[b]{A.~González-Arroyo}
\author*[c]{I.~Soler}

\affiliation[a]{University of Jena, Institute for Theoretical Physics,\\
Max-Wien-Platz 1, D-07743 Jena, Germany}

\affiliation[b]{
Departamento de Física Teórica C-15 and \\
Instituto de Física Téorica UAM-CSIC, \\
Universidad Autónoma de Madrid, Cantoblanco, E-28049 Madrid, Spain}

\affiliation[c]{Dipartimento di Fisica dell’Universit`a di Pisa and INFN, Largo Pontecorvo 3, I-56127 Pisa, Italy}

\emailAdd{georg.bergner@uni-jena.de}
\emailAdd{A.gonzalez-arroyo@pm.me }
\emailAdd{ivan.soler@df.unipi.it}

\abstract{We report results obtained for SU(2) Yang-Mills theory on a four dimensional torus with two directions much smaller than the other two. The small 2-torus is equipped with twisted boundary conditions. This construction provides a way to interpolate from a region in which semiclassical methods can be applied (for small 2-torus size) to the standard infinite volume case. Our simulations at small torus sizes show how the topological charge and the string tension result from a gas of vortex-like fractional instantons. As the size becomes larger the density increases and the separation of structures tends to a constant in agreement with the fractional instanton liquid model picture of the Yang-Mills vacuum.
}

\FullConference{The 41st International Symposium on Lattice Field Theory (LATTICE2024)\\
 28 July - 3 August 2024\\
Liverpool, UK\\}

\begin{document}
\maketitle

\section{Introduction}
In these talks we are presenting some preliminary results of an ongoing project aimed at understanding 
the structure of the vacuum of Yang-Mills theory and its properties such as Confinement. We use a 
modification of the strategy put forward many years ago in a paper by L\"uscher~\cite{Luscher:1982uv}, 
which proposed to use the volume of space as an interpolating parameter between a weak coupling regime 
and the full non-perturbative regime. This led to a series of papers
in which this idea was applied to Yang-Mills theory~\cite{Luscher:1982ma, Luscher:1983gm, vanBaal:1986cw, Koller:1987yk, Koller:1987fq}.
The connection between finite-infinite volume quickly recalls  the idea of volume independence in the large
$N$ limit~\cite{Eguchi:1982nm}. As it is well-known, one of the ways to accomplish the condition of unbroken
center-symmetry is to use twisted boundary conditions~\cite{GonzalezArroyo:1982hz}. This motivated one 
of the present authors
to carry on a program exploring the evolution of the dynamics of the theory formulated on a spatial torus
with twisted boundary conditions as a function of the
volume~\cite{GonzalezArroyo:1987ycm,Daniel:1989kj,
Daniel:1990iz}. It  soon became clear that one has
to go beyond perturbation theory to recover center symmetry. A
semiclassical analysis is necessary~\cite{GarciaPerez:1993jw},
and a leading role in it is played  by a fractional instanton
configuration~\cite{GarciaPerez:1989gt,
GarciaPerez:1992fj} tunnelling among
the two disconnected classical vacua and acting as a kind of kink-solution. This configuration is then responsible
for the building up of the string tension~\cite{RTN:1993ilw}. 
This gas of fractional-kink solutions grows in density as the volume
grows and brings the string tension  to a value close to the infinite
volume one. This  led  us to  propose
the Fractional Instanton Liquid Model~\cite{GonzalezArroyo:1995zy,
GonzalezArroyo:1995ex} as a simple description of the Yang-Mills vacuum
capable of explaining and relating many of its well-known properties.
We refer the reader to a recent review
which includes a more complete list of references~\cite{Gonzalez-Arroyo:2023kqv}.

More recently and also triggered by the goal of understanding the
non-perturbative properties of various theories in terms of
semiclassical ideas various authors followed a similar line of thought
but starting with a complementary situation in which there is a single
compact direction in infinite space~\cite{Poppitz:2012nz,
Poppitz:2012sw}. Different ideas have been
explored to avoid the system to develop a phase transition as one
evolves from small to large sizes~\cite{Bergner:2018unx,
Unsal:2008ch,Shifman:2008ja} (For a recent review and a more complete set of references please consult
Ref.~\cite{Poppitz:2021cxe}).

In this paper, we explore an intermediate path in which two of the four
space-time directions are kept small and two large. The two short
directions are compatified on a torus with twisted boundary conditions. 
This intermediate path has been advocated in several
works~\cite{Gonzalez-Arroyo:2023kqv, Tanizaki:2022ngt,Hayashi:2024qkm}
as a way to bridge the gap among the two strategies. Furthermore,
the string tension appears here as an area law behaviour for large
Wilson loops, instead of as the exponential decay of the correlation
of spatial Polyakov loops as in Ref.~\cite{RTN:1993ilw}.

Having laid down the motivation, we summarise the general strategy and 
ultimate goals of our work. By Monte Carlo generating a large amount  of
SU(2) Yang-Mills configurations on lattices of various sizes and values 
of $\beta$ we have intended to understand how the dynamics depends on
the physical size of the small torus. At small sizes, because of
asymptotic freedom,  one expects that the semiclassical approximation
is capable of describing the system. Checking this is part of our goal.
Ultimately, we would like to understand if there is any relation
between the interpretation of the results in this semiclassical domain 
and the behaviour of the system at large sizes. The ultimate goal is very
challenging but also very appealing.

\section{The semiclassical analysis and the role of Vortex-like fractional instantons}
The system that we are studying in the continuum is pure SU(2) gluodynamics on a $T_2\times R^2$ geometry. We put in a non-zero 't Hooft flux in this torus. Replacing the plane $R^2$ with a large torus has no effect on any of the local observables. Thus, the only relevant external scale present is the linear size $l_s$ of the 2-torus. 
When this size is small, the effective coupling constant is small due to asymptotic freedom. In this regime weak-coupling
methods become reliable. These not only include perturbation theory
but also perturbations around non-trivial local solutions of the
classical equations of motion. In this geometry there are solutions having zero-action that ensure a vanishing trace of the short Polyakov loops. This is the reason why twisted boundary conditions are needed. The leading non-trivial classical solutions  in this setup are SU(2) self-dual configurations
having fractional topological charge $Q=1/2$, that we call vortex-like
fractional instantons. These objects were discovered in
Ref.~\cite{GonzalezArroyo:1998ez} and their main properties studied.
Their size is fixed by the size of the small 2-torus, becoming
independent of the large torus size. Thus, within this large torus,
they appear as circular invariant localised solutions and their
action density decay exponentially for large radial separations. 
Furthermore, if one computes the value of the Wilson loop living in
this big torus and surrounding the center, its value tends to $-1$ for
large loop sizes. Thus, they actually behave as center vortices which
nevertheless have a core which is non-abelian and has a finite
extension.   

\begin{figure}[h!]
        \includegraphics[width=0.45\textwidth]{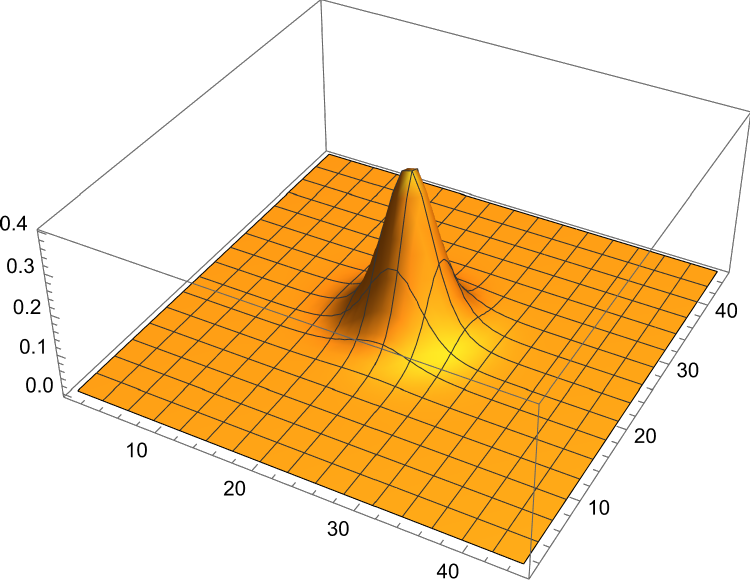}
        \includegraphics[width=0.45\textwidth]{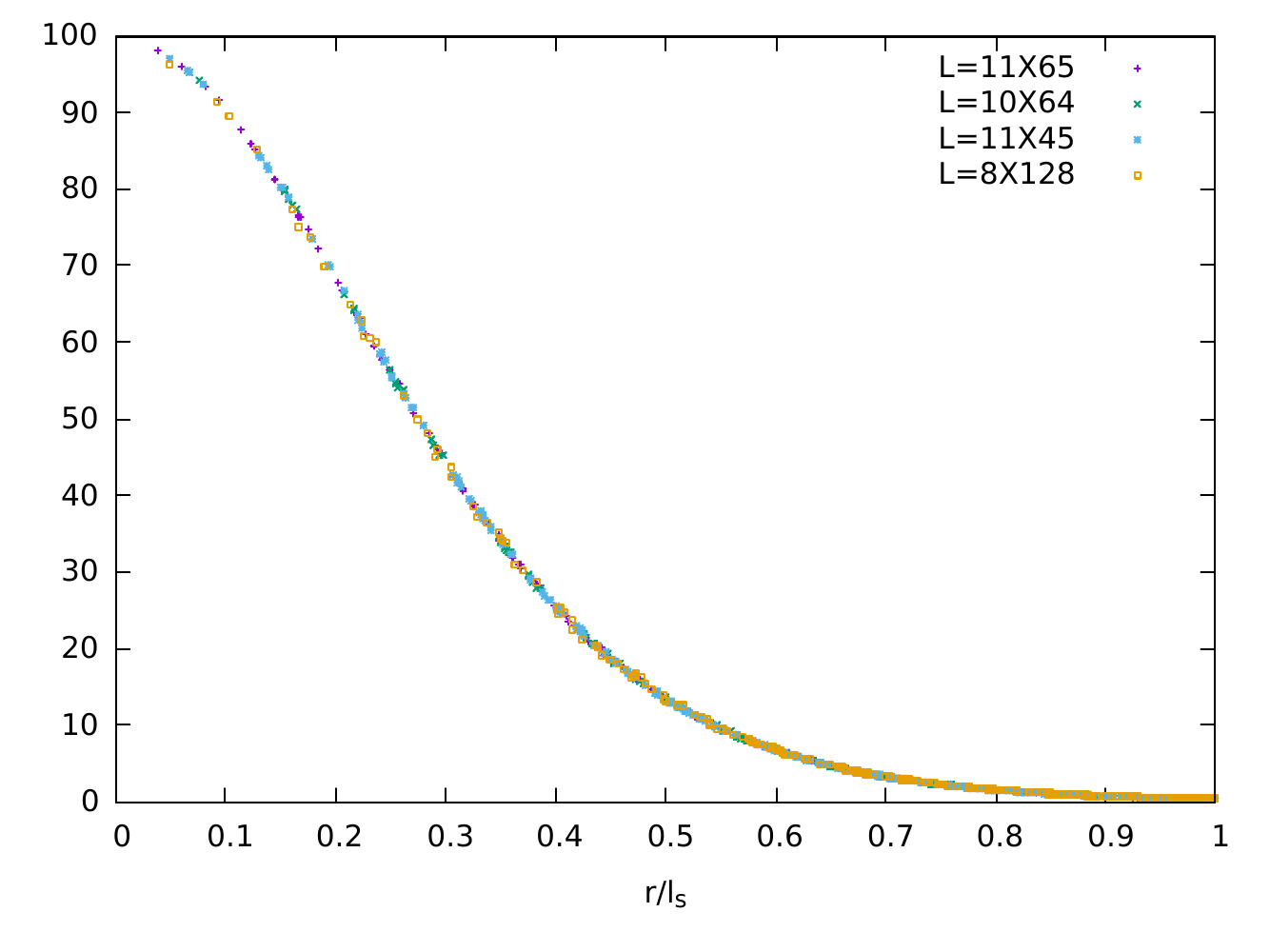}
        \caption{(Left) Vortex-like fractional instanton from a smooth configuration.
        (Right) Rescaled radial profile for different sizes of the torus $N_s$.}
        \label{fig:semiclass_fractional}
 \end{figure}

The solution can be  obtained by minimising the action (as done in Ref.~\cite{GonzalezArroyo:1998ez})
for a similar geometry but with twisted boundary conditions in both
the large and the small torus. The twist in the two orthogonal planes guarantees
that the solution has $Q=1/2$. One can use a lattice gauge theory
approximation on a lattice of size $\Ns^2\times \Nt^2$ with $\Ns\ll \Nt$, 
and a minimisation method as gradient flow or cooling to obtain the
solution with very high precision. The large torus size $\Nt$ becomes
irrelevant provided it is large enough, but the small lattice torus
size $\Ns$ acts as the inverse of a lattice spacing and the distances
have to be divided by it. These properties are clearly seen in
Fig.\ref{fig:semiclass_fractional} where we display the action density
of the solution integrated over the small torus directions. In the
left figure, we see this quantity displayed in a 3D plot as a function
of the position in the big plane. The right figure shows this
integrated action density plotted as a function of the radial
separation to its center, and a very slight rescaling to correct the total integral. A universal figure emerges independently 
of the lattice size employed with lattice errors which are of the
order of the point size.

Although the solution has been obtained by putting a twist also in the 
big torus, solutions having an even number of fractional instantons
also occur if we set periodic boundary conditions in the big plane.
Given that the fractional instanton shape is exponentially localised
around its center, its shape is very slightly affected by the
presence of other fractional instantons at distances $d>l_s$. In
addition, single peaked $Q=1$ instantons can appear that tend to the
shape of ordinary BPST solutions for sizes $\rho\ll l_s$.

The semiclassical approximation predicts that at small torus sizes 
the path integral can be described as a two-dimensional dilute gas of
vortex-like fractional instantons of the described type. Furthermore,
both fractional ($Q=1/2)$ and anti-fractional ($Q=-1/2$) instantons can
appear. Obviously, these are not anymore exact solutions of the
classical equations of motion, but because of the exponential decay of
the action density they will be generated with almost the same
frequency irrespective of whether the local structures have positive or
negative topological charge. 

Indeed,  the semiclassical approximation can also predict how the density 
depends on the torus size. In practice, we are limited because of the 
absence of an analytical formula that describes the solution. 
At the classical level, the probability of
producing a fractional instanton is proportional to the Boltzmann factor.
On the lattice, this implies that this probability should be  proportional to
$\exp\{-\pi^2 \beta\}$ up to finite lattice corrections. Quantum
fluctuations induce an additional weight to each solution and in the
continuum limit, this should combine to give a density which depends
only on the size (in physical units) of the small 2-torus. 

For large torus sizes, the density grows and the gas ceases to be
dilute. Eventually, the density and all the intensive properties become 
independent of the torus size or the boundary conditions, and isotropy
is recovered. It is much harder to predict what happens in this
situation and whether a semiclassical description is at all possible. 
In the long run, our work is intending to understand this transition.
The Fractional Instanton Liquid Model is a proposal of what would
happen in the large size case. The idea is that fractional instanton
solutions continue to populate the vacuum, but their size is no longer
determined by the size of the torus. Since fractional instantons
cannot exist isolated, one expects that the size is precisely dictated
by the distance to the neighbours. That is why the word liquid is more
appropriate than gas. Indeed, configurations with multiple size
fractional instantons can appear on arbitrarily large torus. But 
being able to clarify this or other proposals remains a challenge.

\begin{figure}[h!]
 \centering
        \includegraphics[width=0.45\textwidth]{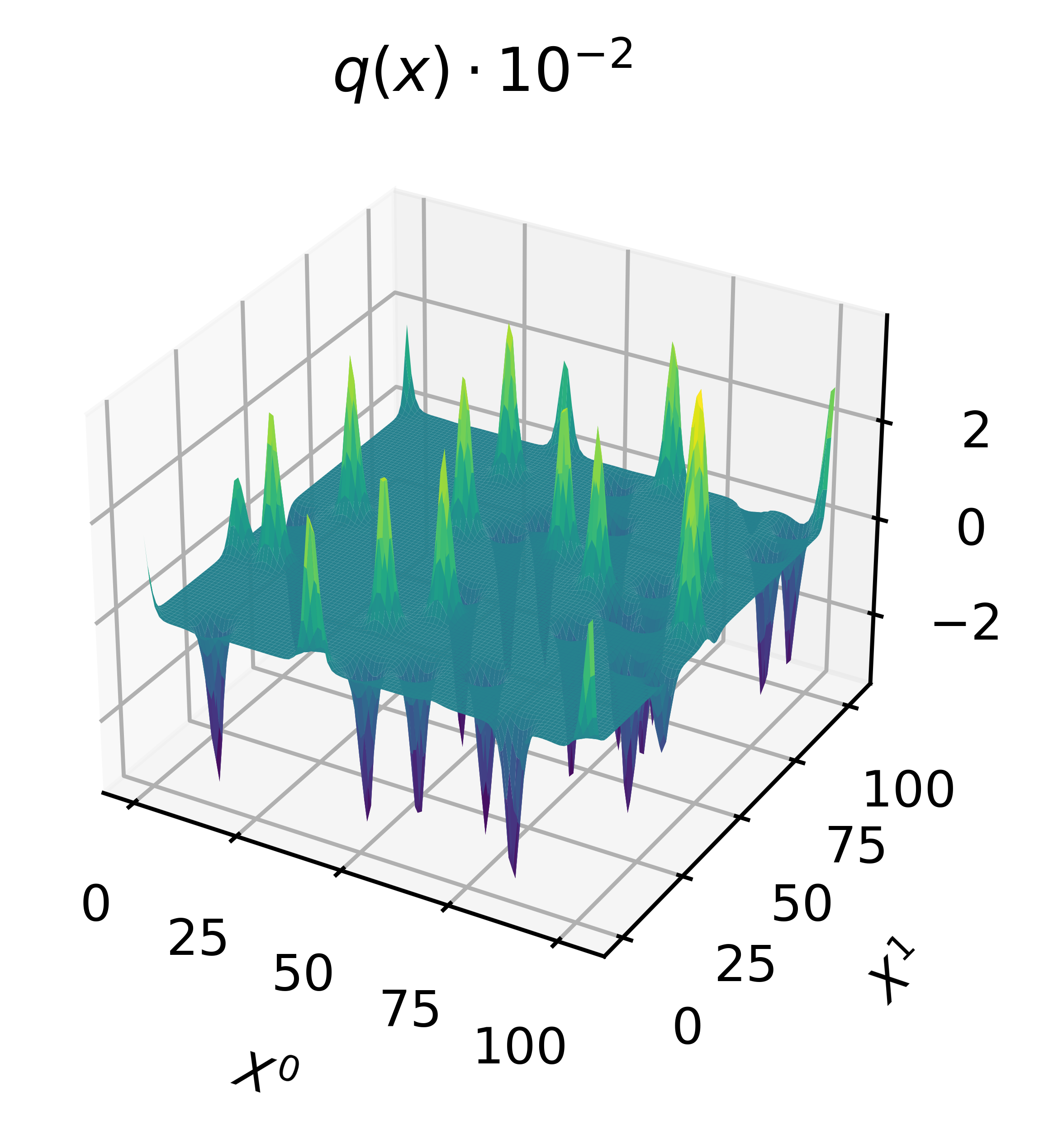}
        \caption{Topological charge density integrated over the small torus for 
        a flowed Monte-Carlo configuration at $\beta=2.6$, $N_s=6$. The flow reveals a configuration with 
        a large amount of fractional instantons.}
    \label{fig:example_conf}
 \end{figure}

Our results at small torus sizes are much less speculative and confirm
our expectations. Indeed, \Figref{fig:example_conf}
shows a typical Monte Carlo generated configuration. It depicts the
topological charge density integrated over the small torus and
displayed as a function of the large torus coordinates. The positive 
and negative spikes have the height and size of the vortex-like 
fractional instantons. Their separation is clearly larger than their size, 
and determining the density is a simple matter of counting the peaks. 

Our work has been that of Monte Carlo generating SU(2) Yang-Mills configurations with Wilson action
on lattices of various sizes  $\Ns^2\times \Nt^2$ with $\Ns\ll \Nt$ and with twisted boundary conditions 
in the small torus and periodic in the large one. We have performed these simulations at many values 
of $\beta$, appropriately tuned to explore a region of physical sizes. As we will show later, this allows
 us not only to check the lattice semiclassical predictions but also to see the corresponding ones in 
the continuum limit. 			       

An important challenge is that of extracting information of the content of the Monte Carlo generated lattice 
configurations. As is well known, these configurations are dominated by short wavelength noise and one needs
some way of filtering out this noise before the large size structures can be revealed. We recently tested one 
method based on adjoint quasi-zero modes~\cite{Gonzalez-Arroyo:2005fzm,GarciaPerez:2011tx,Bergner:2024njc} but it turns out to be too costly 
to be implemented in an exploratory study. Thus, we have judiciously used gradient flow as a filtering method. Indeed, \Figref{fig:example_conf}  
has been obtained after flow. In addition to flow, one has still to scan the resulting configurations and identify 
the presence of fractional instantons and ordinary instantons in it. This seems easy in the example configuration displayed, 
but becomes rather challenging when the density is high. 
The procedure that has been used is described in the next section.

\section{Identification}
The identification of topological objects in lattice configurations is important to test the validity of instanton
models for the vacuum of Yang-Mills theories. For our identification, we followed a two-step procedure; first, a certain 
amount of gradient flow is applied to smoothen out the configurations and filter the UV noise to reveal the IR structures associated with topological objects (instantons and fractional instantons). Then, we scan the topological charge density to find which of the remaining structures indeed corresponds to fractional or integer instantons. The main advantage of our study is that we can slowly study the system as a function of the torus size $l_s$ and monitor the objects as their density increases.

\begin{figure}[h!]
 	 \includegraphics[width=0.45\textwidth]{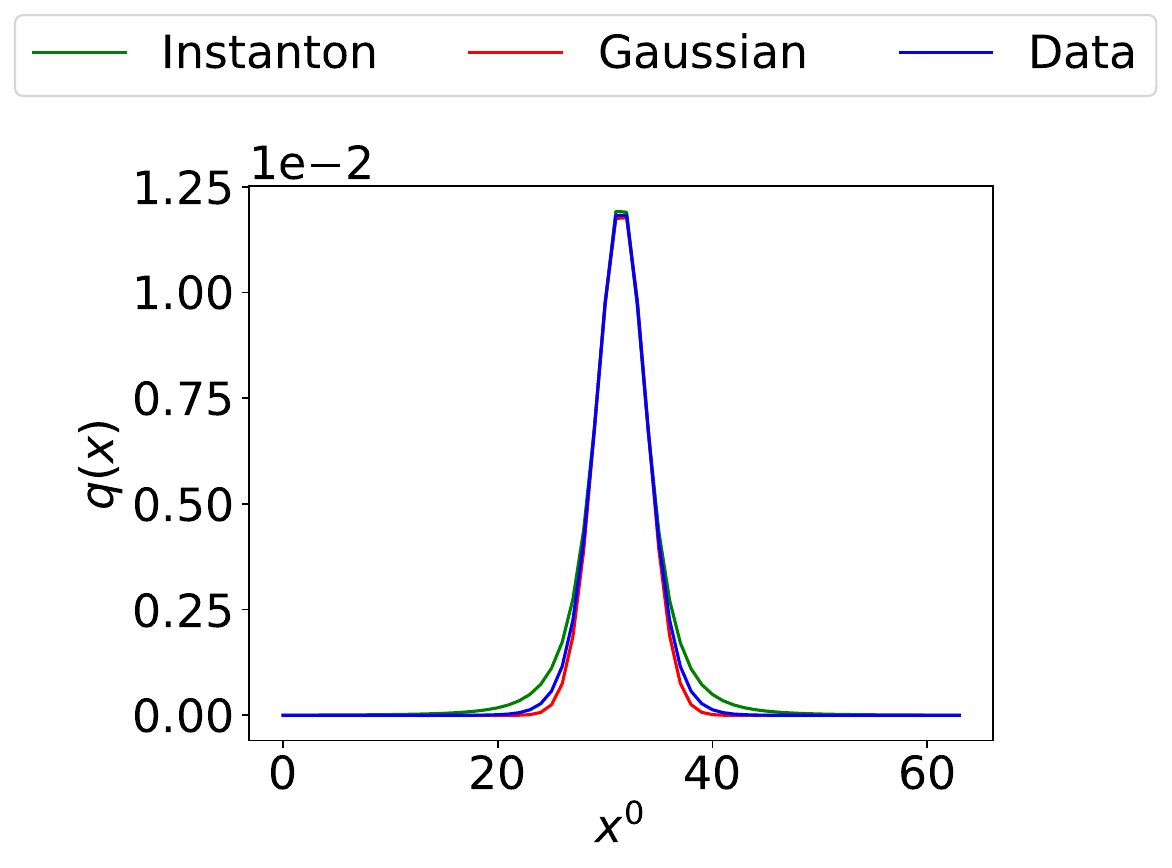}
 	 \hspace{1cm}
	\includegraphics[width=0.33\textwidth]{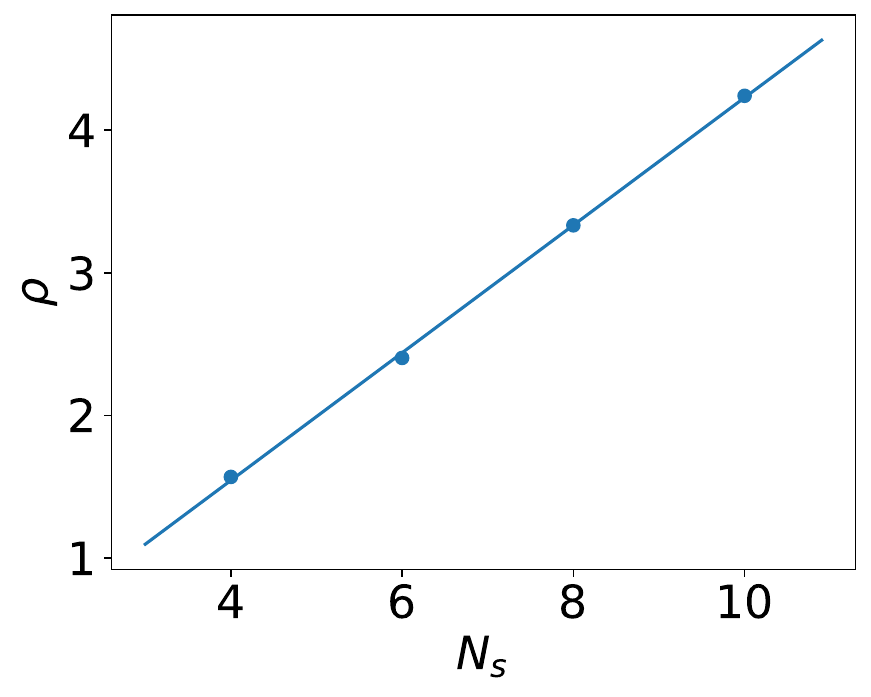}
 	  \caption{ (Left) 1D section of the topological density $q(x)$ of a smooth fractional instanton. (Right) $\rho$ of smooth fractional instantons obtained at different torus sizes.}
 	  \label{fig:smooth_fit}
 \end{figure}

To identify which IR structures should be considered, we first sum the topological density over the short $T_2$ directions, this is a good approximation as long as only one object fits in the small 2-torus. Then, we find all the local maxima (and minima) in the topological charge density, and we perform a fit to the BPST solution using the peak and its nearest neighbour points
\begin{align}
	q(x,y)=Q \frac{\rho^2}{(r^2+\rho^2)^2},
\end{align}
 obtaining for each peak the normalization factor $Q$ and the size $\rho$. Notice that the BPST fit seems to accurately describe the object at least around the peak \Figref{fig:smooth_fit}. At the tails, however, the fractional instantons decay exponentially \cite{GonzalezArroyo:1998ez}, and so the fit fails. Nevertheless, because we fit just around the maxima, the local properties do not depend so much on the tails; in fact, we tested the fit for fractional instantons at different torus sizes and obtained the expected linear scaling of $\rho$.

\begin{figure}[h!]
\includegraphics[width=0.45\textwidth]{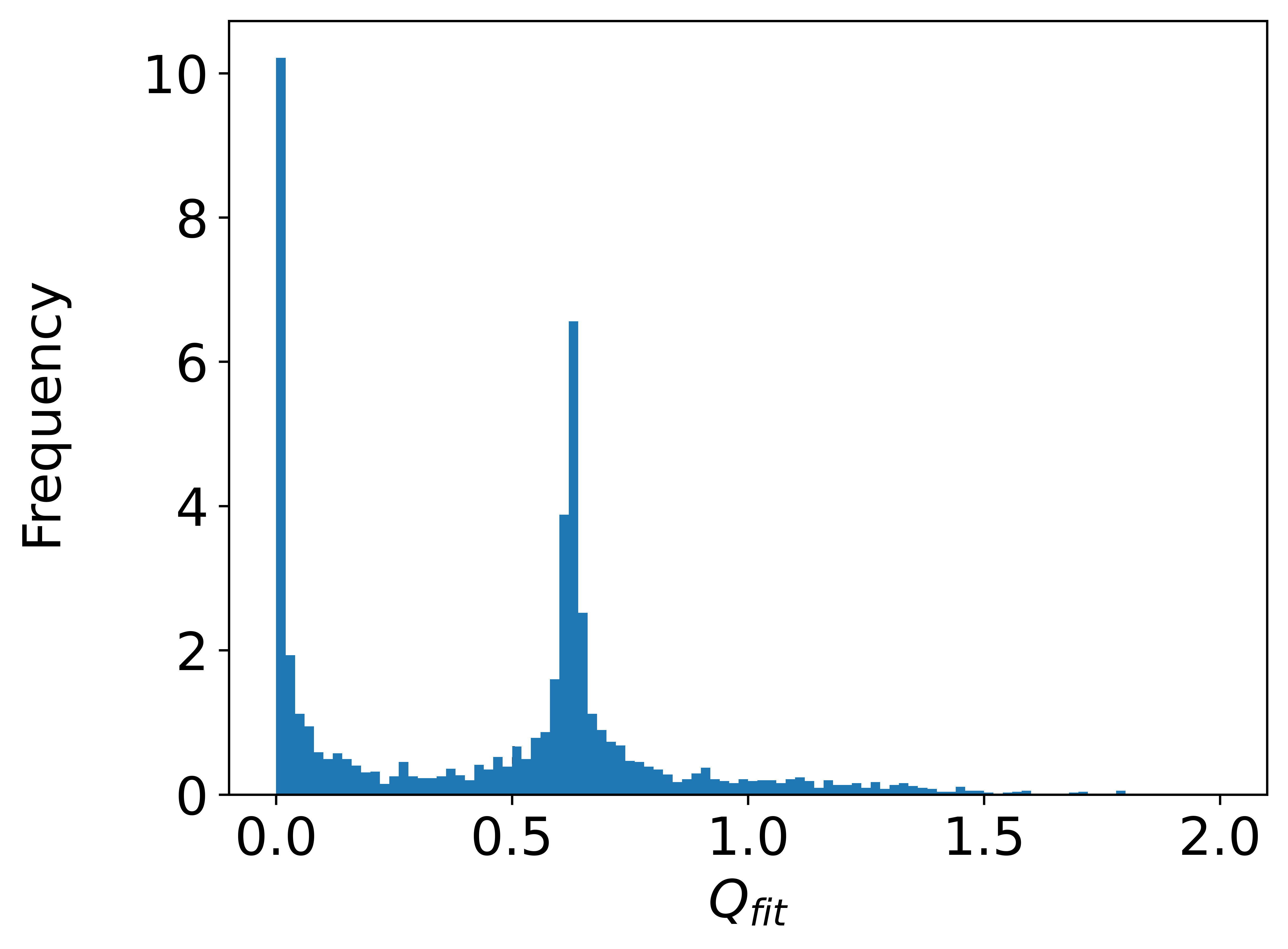}
 	\includegraphics[width=0.45\textwidth]{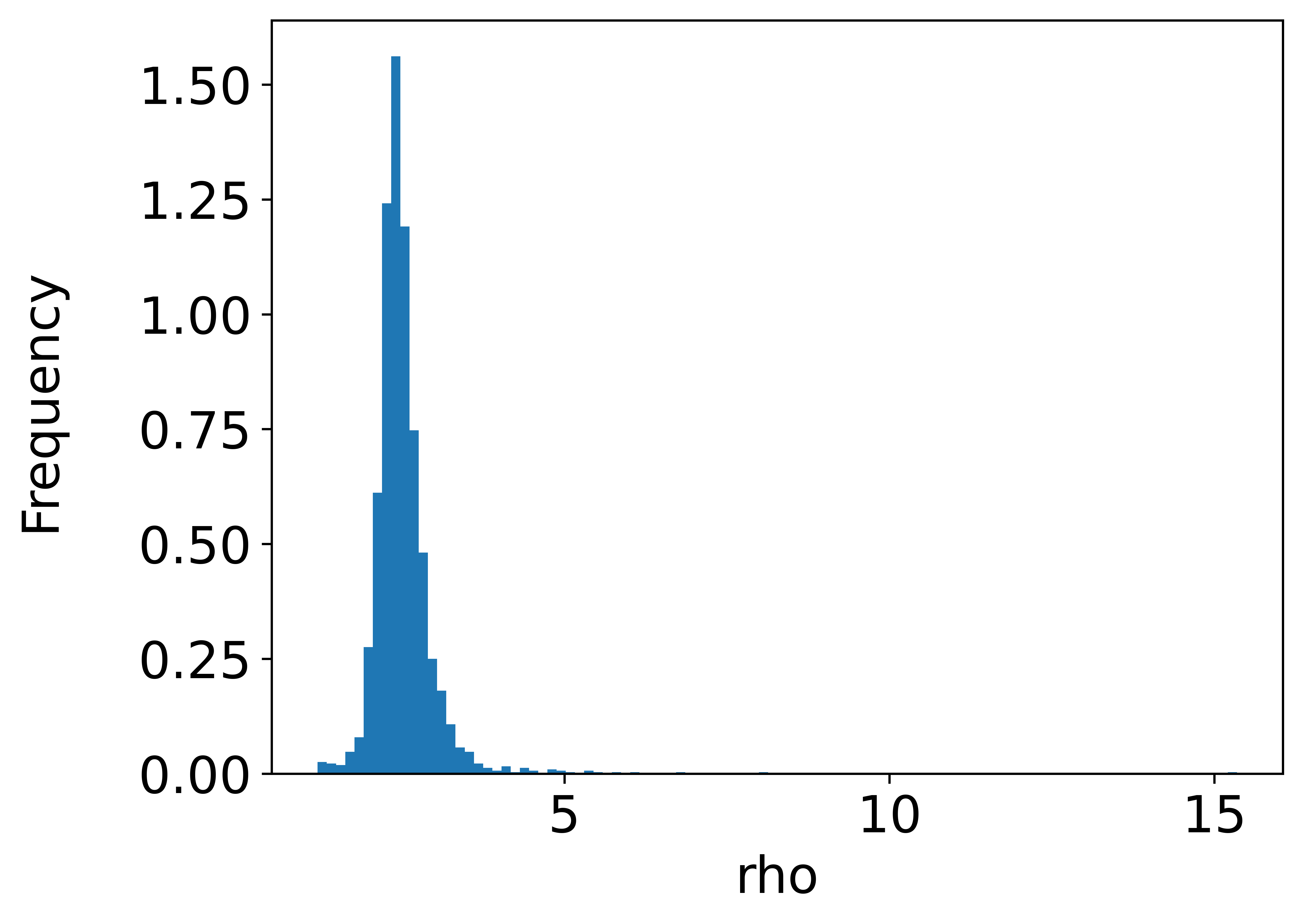}
 \caption{Histograms for $Q_{fit}$ and $\rho$ of the population of local maxima. Simulation was performed at $\beta=2.6$, $N_s=6$, $t_{gf}=15$.}
 \label{fig:histograms}
 \end{figure}

  When scanning all the peaks in Monte-Carlo configurations, one obtains a population of objects with different parameters. The algorithm for the identification is the following: we first discard those peaks whose fitted maxima lie at a distance less than one lattice spacing with each other, and we also discard peaks with a size smaller
 than one lattice spacing $\rho < 1$. The histogram on the normalisation $Q$ is clearly peaked at $Q_{FI}$ corresponding to the smooth fractional instantons \Figref{fig:histograms}, and similarly, a peak in the $\rho$ histogram is obtained. Therefore, we take as fractional instanton those peaks which have a fitted value of $Q_{peak}(1-\lambda) < Q_{fit} < Q_{peak}(1+\lambda)$ where we took $\lambda=0.75\pm0.1$, which we varied to estimate how sensitive the identification is to the parameter $\lambda$. 
 
\section{Properties of fractional instantons as a function of $l_s$}
\begin{figure}[h!]
 	\includegraphics[width=0.45\textwidth]{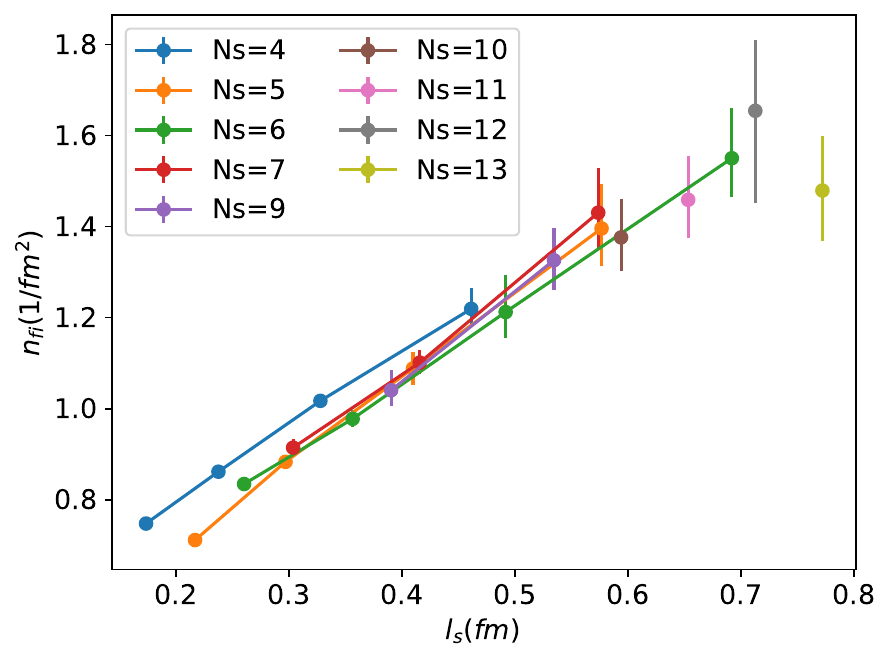}
 	\includegraphics[width=0.45\textwidth]{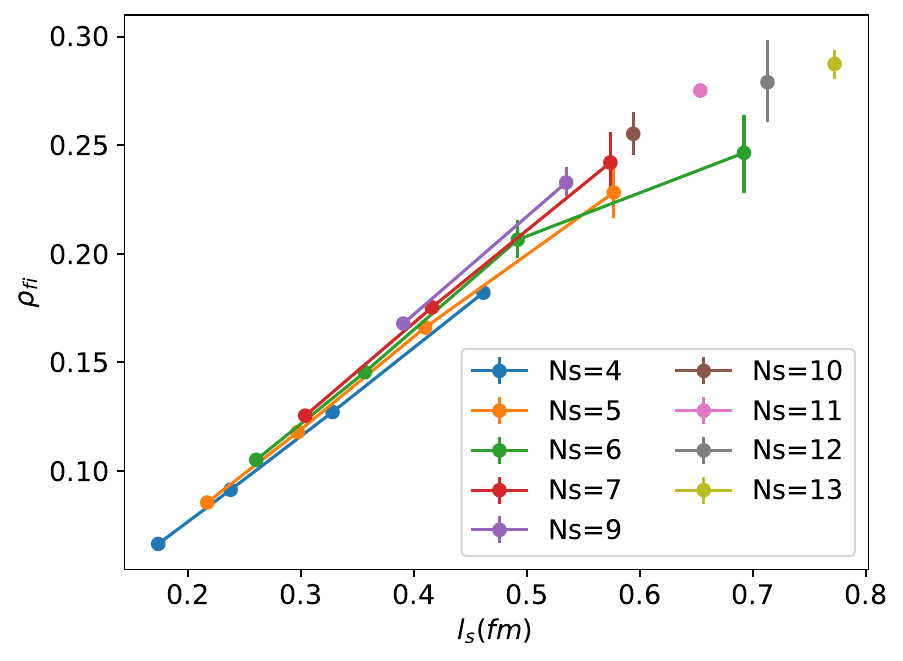}
 	\caption{(Lef) Density of fractional instantons $n_{fi}$ as a function of $l_s$. (Right) Mean value of the size of the fractional instantons $\rho$ as a function of $l_s$. }
 	\label{fig:density}
 \end{figure}

 We applied the identification algorithm defined in the previous section to $T_2\times R^2$ Monte-Carlo simulations at different $l_s$. The gradient flow time $t_{gf}$ (in lattice units) was tuned such that all the ensembles had the same smoothing radius $\tau=\sqrt{8t_{gf}}\;a$ in physical units\footnote{We use fermi as a physical unit, defined by declaring that the infinite volume string tension is 5 fm$^{-2}$}. We computed the density of fractional instantons as a function of $l_s$ and plotted the results in \Figref{fig:density}. From this figure, we can see how the density of fractional instantons increases with $l_s$ as predicted by semiclassics and also how, at large $l_s$, the error bars are broader due to the identification becoming harder as the density and the fluctuations increase. Nevertheless, we obtained a very consistent scaling up to $l_s=0.7 \text{fm}$; only the ensemble with the smallest size in lattice units $N_s=4$ departs from the trend, which we expect to be related to strong finite volume corrections to the fractional instanton solution (similar behaviour was obtained for test $N_s=3$ configurations). Furthermore, the mean value of the size of the fractional instantons follows the behaviour previously seen with the smooth configurations, suggesting that at $l_s<0.7\text{fm}$, we are still in the semiclassical regime. As a final remark, notice that the evolution from small to large $l_s$ is rather smooth and no trace of phase transition is observed. This is in line with the expectation that twist protects the system from center symmetry breaking at small volumes. Of course, a more systematic study of the Polyakov loop is needed to make this statement more precise.
 \begin{table}[h!] 
	 \begin{center}
 	\begin{tabular}{ |c|c|c| } 
 	\hline
 	$ \beta $ & a & $t_{gf}$\\
 	\hline
 	2.4 &  0.11530  & 3.52\\
 	\hline
 	2.5 & 0.08194 &  7.34 \\
 	\hline
 	2.6 & 0.05938 & 15 \\
 	\hline
 	2.7 & 0.04337  & 27.37\\
 	\hline
 \end{tabular}
 \hspace{1cm}
 \begin{tabular}{ |c|c|c| c | } 
 	\hline
 	$N_s$ & $N_r$ & $\tau=\sqrt{8t_{gf}}\;a$ & twist \\
 	\hline
 	4-13 & 64, 128 & 0.650 fm & (2,3) \\
 	\hline
 \end{tabular} 
 \end{center}
 \caption{Parameters used for the Monte-Carlo generated configurations.}
 \label{tab:parameters}
 \end{table}

\subsection{Semiclassical regime}
 In the semiclassical regime, one expects perturbative estimates to be a good approximation in computing the quantum weight. Unfortunately, there is no simple analytic formula for the fractional instantons to allow such computations. However, the spirit of the calculation is similar to the one derived by 't Hooft for $Q=1$ BPST instantons \cite{tHooft:1976snw}.
 
 Let us focus on a dimensionless quantity that we call diluteness $D$ and defined as follows
 \begin{align}
     D(\beta,\Ns)=(N_{FI}+N_{AFI})\Ns^2/\Nt^2,
 \end{align}
 Since the probability of producing a fractional instanton is proportional to its Boltzmann weight, one expects the  following $\beta$ dependence for the diluteness
\begin{align}
    D=\mathcal{A}(\Ns)\beta^2\text{exp}(-\beta\pi^2).
    \label{eq:dilute}
\end{align}
The leading dependence is the exponential one with an exponent proportional to the classical action of the fractional instanton. The $\beta^2$ comes from the $g$ contribution of each zero mode to the determinant of the perturbations around the solution\footnote{In the case of $SU(2)$ instantons one finds eight zero modes, leading to a $\beta^4$ dependence in \cite{tHooft:1976snw}}. We collect in $\mathcal{A}(\Ns)$ the factors only slightly dependent on $\beta$ and coming from the quantum fluctuations around the fractional instanton solution. Incidentally, Eq.~\ref{eq:dilute} is the same formula occurring in the $T_3\times R$ analysis of Ref.~\cite{RTN:1993ilw}.

\begin{figure}[h!]
 	\includegraphics[width=0.49\textwidth]{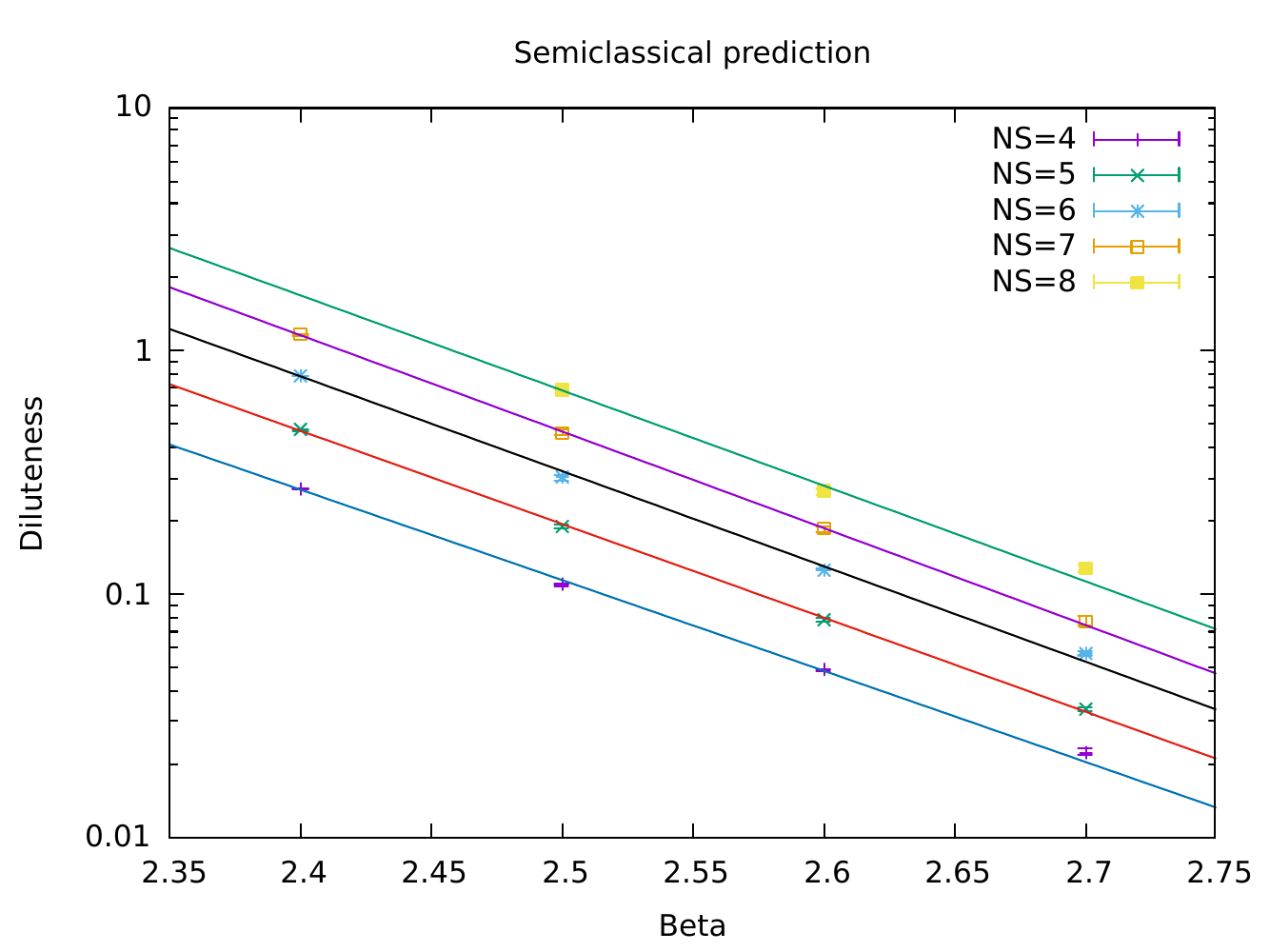}
    \includegraphics[width=0.49\textwidth]{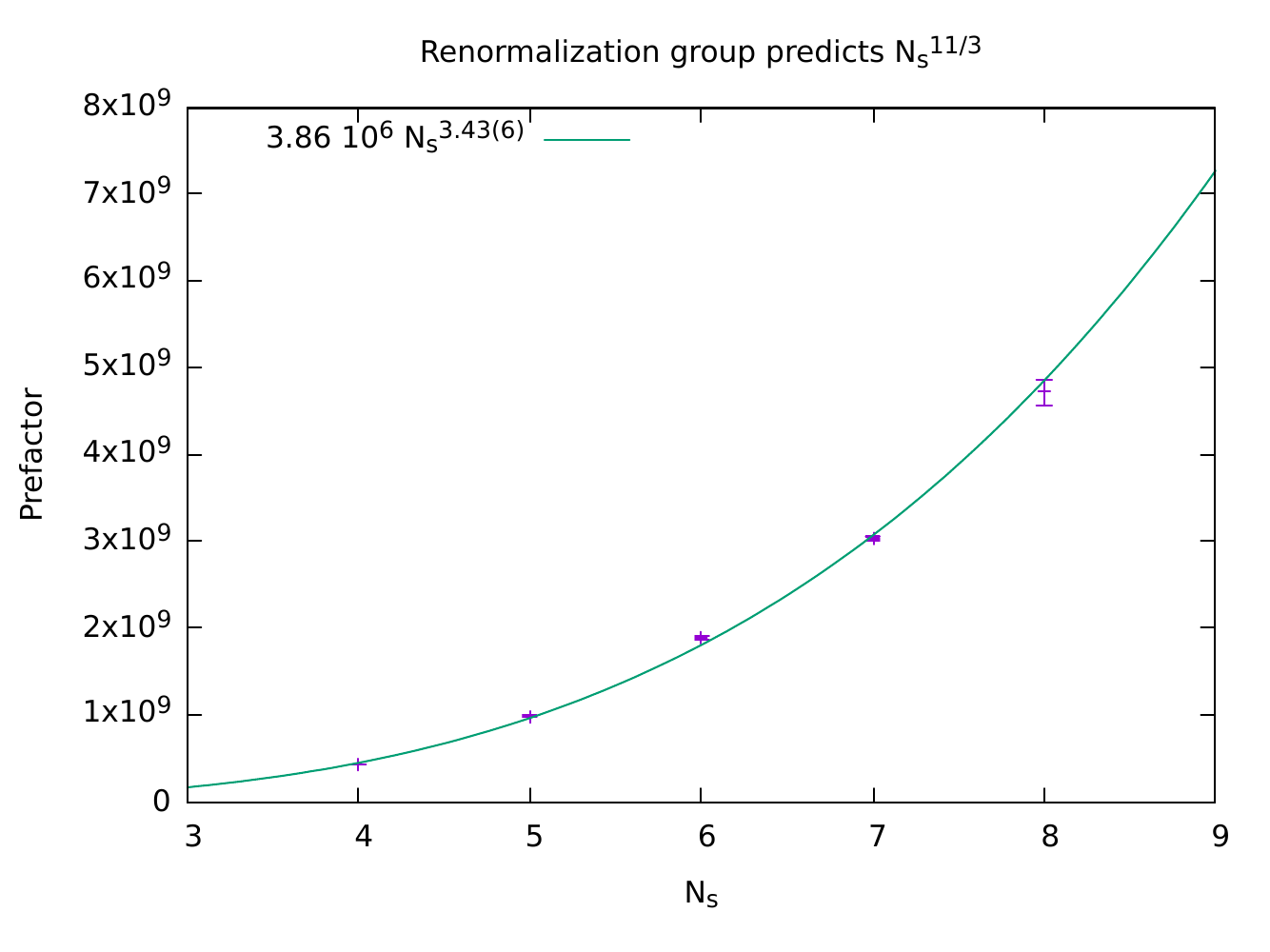}
 	
 	\caption{(Left) Diluteness measured on our Monte-Carlo configurations and the semiclassical prediction.(Right) Fitted prefactor $\mathcal{A}$ as a function of $N_s$, the line shows the fit to the semiclassical prediction.}
 	\label{fig:semiclassical}
 \end{figure}

We have plotted our results for the diluteness obtained from our identification in log scale in \Figref{fig:semiclassical}. We see that the results follow perfectly the form \Eqref{eq:dilute}, showing how well the semiclassical approximation works in those ensembles. The result is particularly striking as the diluteness is changing by more than one order of magnitude. The fit to \Eqref{eq:dilute} also allows us to extract the $\mathcal{A}$ prefactor. Even though we don't have an analytical expression for this factor, we know the precise dependence with $\Ns$ due to the renormalisation group flow. The diluteness is dimensionless, and so scale independent; therefore, the scale dependence of $\mathcal{A}(l_s/a)$ must cancel that of the coupling $\beta(a)$. At first order in the gauge coupling $g$, and in the continuum limit, one must have $\mathcal{A}\propto \Ns^{4\pi^2b_0}$ where $b_0$ is the one-loop beta function. In particular, for SU(2), we have that $4\pi^2 b_0=11/3\approx 3.66$. The fits we obtained \Figref{fig:semiclassical} show a dependence with $\Ns$ with an exponent $3.43(6)$, which agrees with the theoretical value up to contributions coming from higher orders in perturbation theory and finite-size corrections. These results show that the semiclassical formulas hold for quite a large range of $l_s$ and that the identification algorithm, even though simple, seems to be effective enough.

Let us now focus on the consequences of the semiclassical description of the Yang-Mills vacuum in terms of a two-dimensional gas of fractional instantons. First of all, the distribution has to follow the Poisson law, which we have explicitly tested. A very important consequence is its implication on Confinement. Here, it is crucial that our fractional instanton solutions behave as $Z_2$ vortices at large distances. As is well-known, a 2D gas of these vortices confines. 
Furthermore, a very simple calculation relates 
the value of the string tension to the density of these objects. The calculation is exact for a two-dimensional $Z_2$ gauge theory. 
We call this calculation the  {\em thin abelian vortex approximation} (TAVA). In this simple theory given a Wilson loop of area $A$ its value for  a given configuration equals $(-1)^n$ where $n$ is the number of $Z_2$
vortices contained in the area $A$. 
The Poisson distribution now tells us that the probability of having $n$ vortices inside the area $A$ goes as follows
\begin{align}
 P(n)= \frac{\bar{n}^n}{n!} e^{-\bar{n}}\; ,   
\end{align}
where $\bar{n}$ is the mean number of vortices in the area $A$. This number is given by $\bar{n}=\rho A$, where $\rho$ is the two-dimensional density. Thus, the expectation value of the Wilson loop becomes 
\begin{align}
W(A)=\sum_{n=0}^{\infty} (-1)^n  \frac{(\rho A)^n}{n!} e^{-\rho A} = e^{-2 \rho A}
\end{align}
from which one can read a perfect area law with a string tension given by two times the density. Although the calculation has been performed in the TAVA approximation it is expected to hold for our non-abelian thick vortices in the limit of low densities. The thickness of the object generates a perimeter contribution but will not affect the calculation of the string tension if the probability of two vortices overlapping is very small. 

\begin{figure}[h!]
    \centering
 	\includegraphics[width=0.49\textwidth]{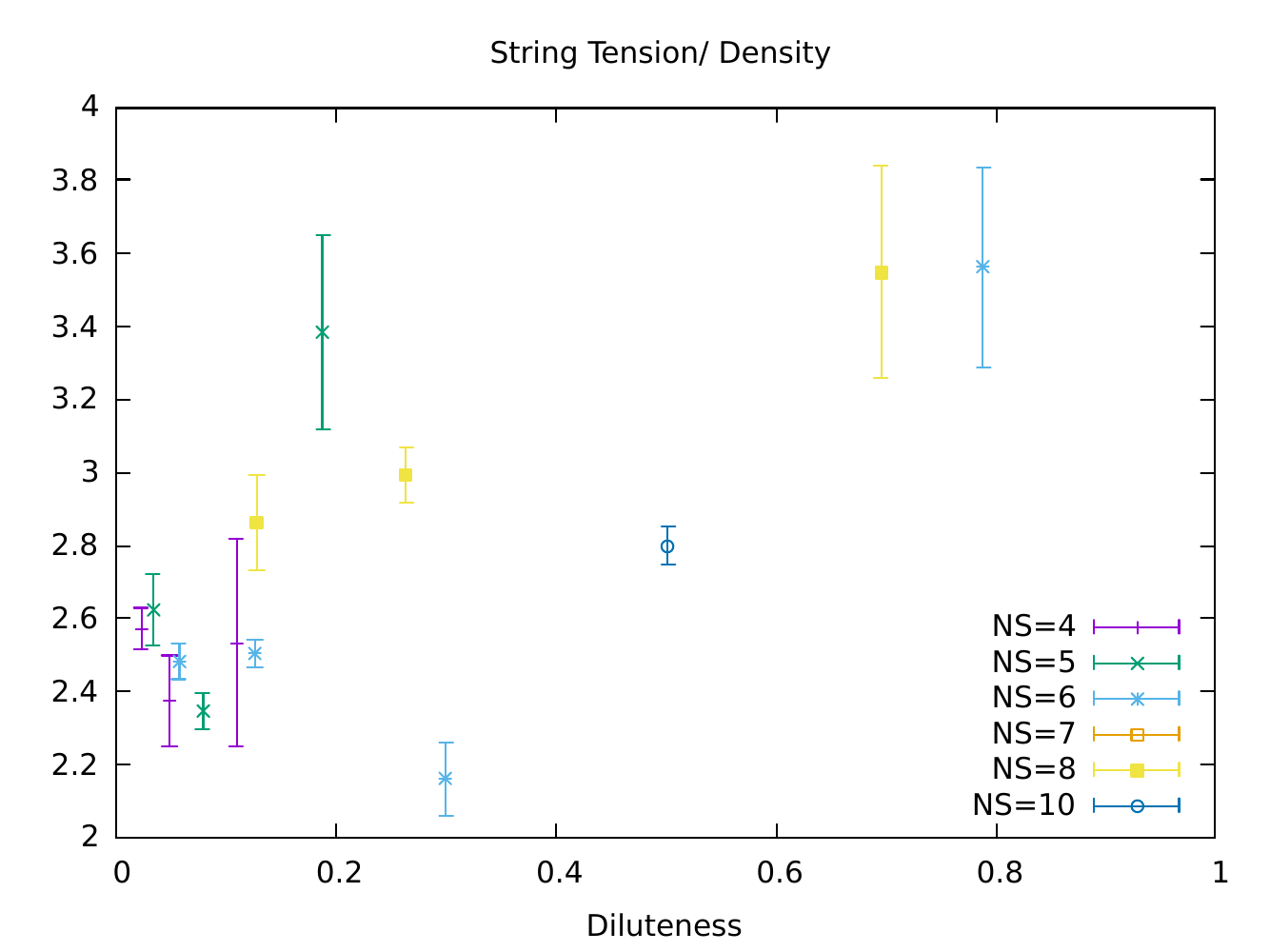}
 	\caption{The string tension over fractional instanton density shows a strong correlation for all the ensembles.}
 	\label{fig:string_tension}
 \end{figure}

 To check whether this relation holds, we have computed the square Creutz ratios $\chi(R)$for our flowed configurations. Previously, we measured their value for the smooth fractional instanton solution, and we found that the ratio peaks at $R\sim l_s$ to a value around  2.3-2.4. For larger values, it drops down to the asymptotic value of 2. 
 Unfortunately, Creutz ratios for large R are very noisy. Thus, we computed an average over Creutz ratios of size above $R=l_s$. The results are displayed in \Figref{fig:string_tension}. Where one can see that at low densities the ratio $\sigma/\rho$ slightly exceeds the asymptotic value. The ratio does not grow much for higher values of the diluteness despite the fact that the density changes by a large amount for the different ensembles. The correlation of both quantities is very clear, but better measurements are needed to test the evolution with the Creutz ratio size.

\subsection{Towards the large volume limit}
All the results so far were focused on the semiclassical region and showed a consistent picture in which the fractional instantons are responsible for the non-vanishing string tension. The growth of the density of fractional instantons goes in parallel with that of the estimates of the string tension (through Creutz ratios) as $l_s$ increases.
However, the growth of both quantities has to stop in order to match the behaviour at infinite volume.
 In particular, the string tension needs to saturate at its physical value, and the system needs to decouple from the size of the torus, keeping only one physical scale $\Lambda$. It is at this point that the system becomes independent from the boundary conditions, and the infinite volume limit is reached.
  
\begin{figure}[h!]
 	\includegraphics[width=0.45\textwidth]{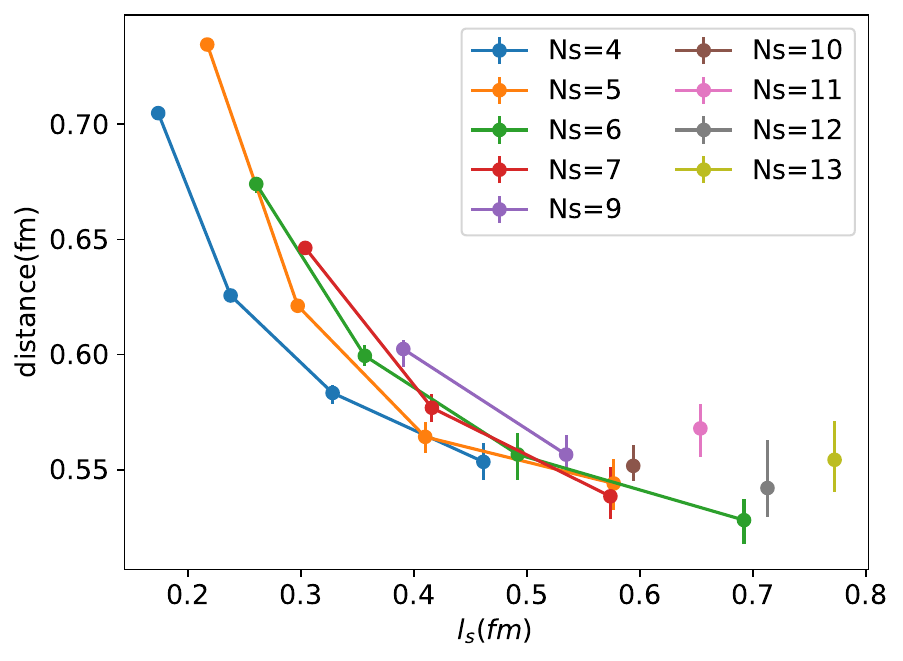}
 	\includegraphics[width=0.43\textwidth]{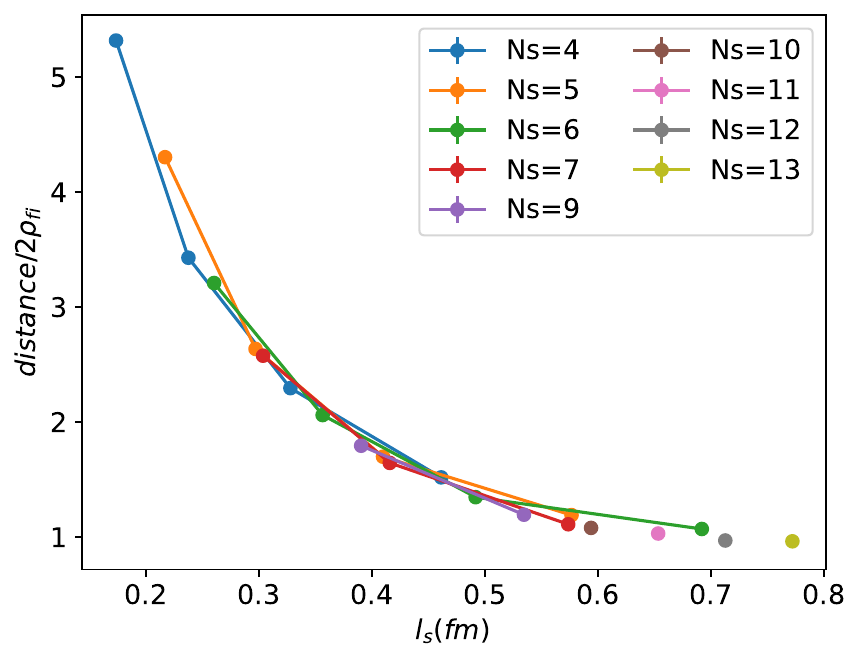}
 	\caption{(Left) Mean distance between nearest fractional instantons as a function of $l_s$. (Right) The same distance divided by two times the mean value of the size of the fractional instantons. The constant behaviour at large $l_s$ is the expected one at large volume.}
 	\label{fig:distance}
 \end{figure}

In the original proposal of the fractional instanton liquid model~\cite{GonzalezArroyo:1995zy,
GonzalezArroyo:1995ex} The mechanism under which the fractional instantons decouple from the torus is when the size reaches a value at which the probability of producing a new fractional instanton is not suppressed anymore. Before that point is reached, the size starts to be driven more by the proximity of nearby structures rather than by the torus size.
Furthemore, as this mean separation becomes smaller than $l_s$ one can have several fractional instantons fitting within the small torus. In this way the liquid becomes four-dimensional and isotropy is recovered.
To see if these ideas match with our results, we computed the distance between nearest fractional instantons neighbours \Figref{fig:distance}, which decreases with $l_s$ as expected and seems to saturate at $l_s\sim 0.7\text{fm} $. When plotting $\rho/(d/2)$ one can clearly see a strong correlation between these quantities and \Figref{fig:distance} shows how the size tends to $\rho\sim d/2$ at large $l_s$, as expected in the large volume limit.  

Another interesting hint towards the infinite volume results is the fact that at large $l_s$, the size of the fractional instantons seems to depart from the semiclassical behaviour \Figref{fig:density}. Strikingly, we checked that around that point is where the string tension seems to saturate to its physical value. These are all hints that $l_s=0.7 \text{fm}$ is the end of the semiclassical regime and that one is approaching large volume results. This is actually in line with the previous results in \cite{GarciaPerez:1993jw,GonzalezArroyo:1995zy,RTN:1993ilw}, obtained with a different lattice geometry $T_3\times R$. It is quite remarkable that even close to the semiclassical edge, the analytical formulas still hold, as seen from \Figref{fig:semiclassical}. One is tempted to push our computations towards larger $l_s$; however, the difficulty increases as the fractional instantons start to populate also the $T_2$ direction and the liquid becomes four-dimensional. Therefore, for larger $l_s$, one cannot integrate over the small torus, making the signal more noisy. We are at present working on developing strategies to be applied to this regime. 
 
 \section{Gradient flow dependence}
\begin{figure}[h!]
 	\includegraphics[width=0.45\textwidth]{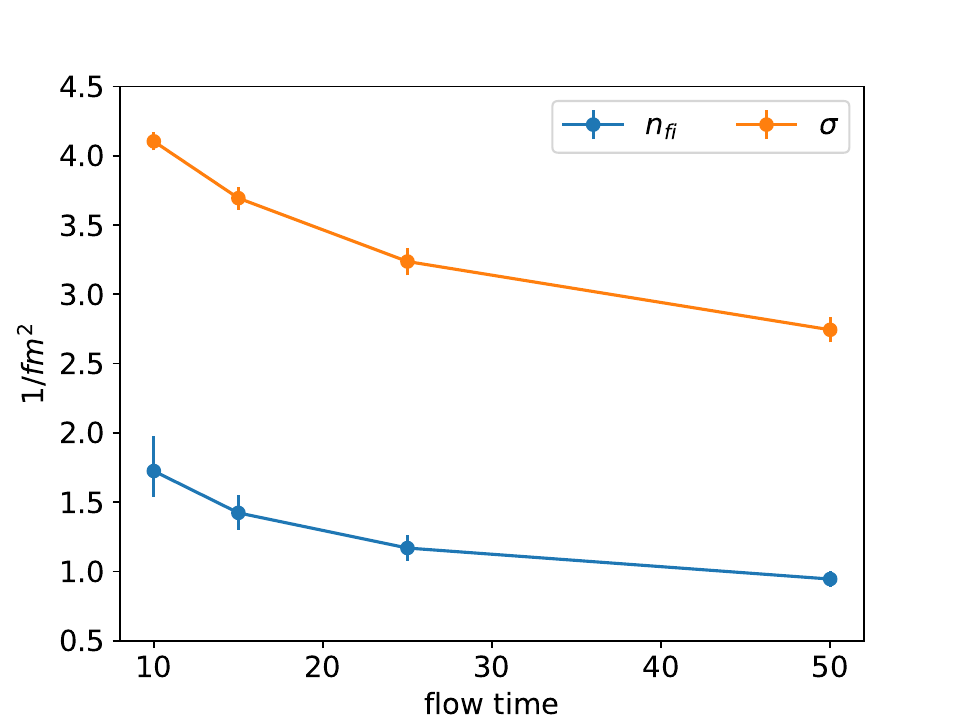}
 	\includegraphics[width=0.45\textwidth]{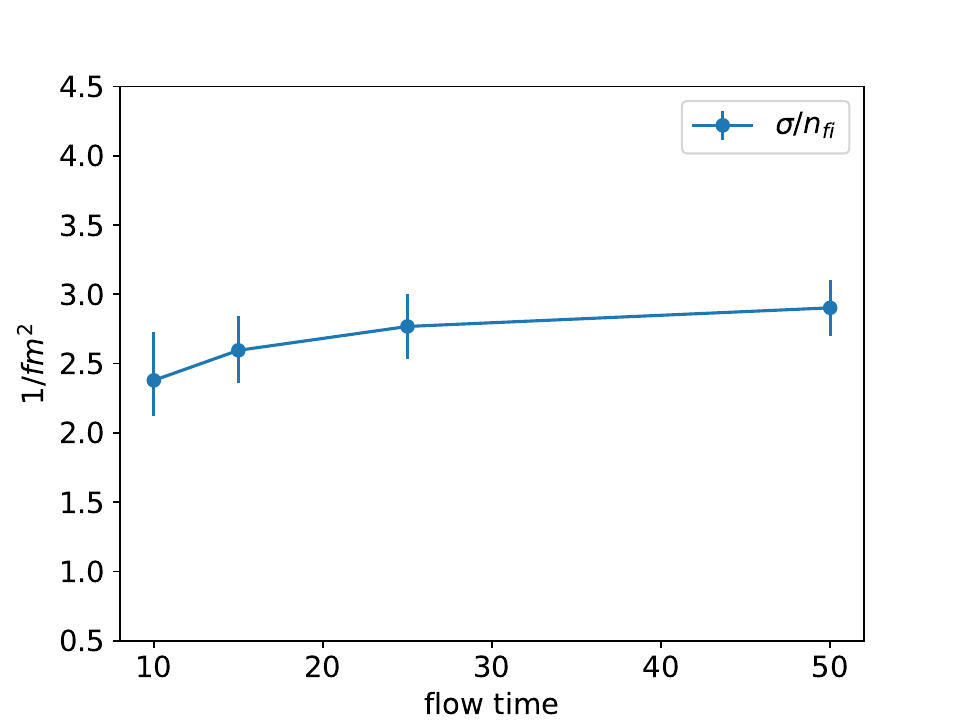}
 	\caption{(Left) Dependendence of the string tension $\sigma$ and the density of fractional instantons $n_{fi}$ as a function of the flow time for an ensemble with $l_s\sim0.6 \text{fm}$. (Right) the ratio $\sigma/n_{fi}$ remains constant during the whole flow.}
 	\label{fig:flow_dependence}
 \end{figure}

 As a final remark, we want to address the dependence of our studies on the gradient flow used to smooth out the gauge configurations. The effect of the gradient flow has been quite studied during the past years, which led to a good understanding of its effects on configurations with non-trivial topological charge. When using Wilson flow, there are mainly two effects that change the density of fractional instantons
 \begin{itemize}
 	\item Fractional instantons can merge together, producing a single $Q=1$ instanton. At that point, the Wilson flow shrinks the instantons, and when $\rho<=1a$, the instanton evaporates as it "falls through the lattice" \cite{GarciaPerez:1993lic, deForcrand:1997esx}.
 	\item Pairs of fractional/anti-fractional instantons annihilate each other, as they are not protected topologically, and the flow tends to minimise the total action in the system.
 \end{itemize}
 Both these effects lead to a decrease in the fractional instanton density with the flow time. While the instanton evaporation is mostly a lattice artifact (not entirely since in the absence of twist there are no instantons on the torus \cite{Braam:1988qk}) that can be compensated using an overimproved action in the flow \cite{GarciaPerez:1993lic, deForcrand:1997esx}, the annihilation is an effect that also happens in the continuum and, in principle, cannot be avoided in the flow. For this exploratory study, we used Wilson flow due to the expensiveness of the overimproved flow. Nevertheless, we expect the correlation of the string tension and the density of fractional instantons to be left invariant under the flow if the fractional liquid model is correct. In fact, when plotting the density and the string tension along the flow for a particular ensemble dense enough ($l_s\sim0.6$ fm) \Figref{fig:flow_dependence}, we can see how both quantities decrease while keeping their ratio constant $\sigma/n_{fi}\approx 2.7$, showing further evidence on the semiclassical predictions and the relation of fractional instantons with Confinement.
 
\section{Conclusions}
In this talk we have presented  preliminary results of 
our investigation on the evolution of the dynamics of SU(2) Yang-Mills
fields on a $T_2\times R^2$ geometry as a function of the small torus
linear size $l_s$. It seems that our use of twisted boundary
conditions avoids the presence of a phase transition when going from
small to large sizes. The small volume regime can be well
described by the predictions of a semiclassical analysis in which the
system appears as a two-dimensional gas of vortex-like fractional
instantons that fully occupy the small torus. We have successfully tested these
expectations both for the lattice and the continuum. Obviously, this
gas gives rise to a non-zero string tension as measured by Wilson
loops in the large plane, and its value is determined by the density
of the gas. As the size gets larger, the density increases and it is
much harder to identify the structures. Nevertheless, we have studied
the region in which the density is high and found that the mean
separation between the objects tends to a constant in physical units
of the order of 0.7 fermi. This is in line with the predictions of the
fractional instanton liquid model and matches with the value which
emerged from the previous $T_3\times R$ studies. We also monitored the
evolution of the string tension in this process and observed that it
saturates to a value close to the one measured at infinite volume and
not too far from the value it gets in the region where the semiclassical
description is still good. It seems hard to believe that a similar
value could arise from a completely different mechanism. However, to
better understand this point, a more thorough study of the transition
region and the approach to infinite volume is necessary and demands a
refinement of our techniques. We are working on it at the moment.

\acknowledgments{ G. B. is funded by the Deutsche Forschungsgemeinschaft (DFG) under Grant No.~432299911. A.G-A acknowledges
support by the Spanish Research Agency (Agencia Estatal de Investigaci\'on) through the grant IFT Centro de Excelencia Severo Ochoa CEX2020-001007-S, funded by \\
MCIN/AEI/10.13039/501100011033, and by grant PID2021-127526NB-I00, funded by \\
MCIN/AEI/10.13039/501100011033 and by “ERDF A way of making Europe”. I. S. acknowledges support from project PRIN 2022 “Emerging gauge theories: critical properties and quantum dynamics” (20227JZKWP) and the support from DFG through the Grant No.~431842497 . Part of the computing time for this project has been provided by the compute cluster ARA of the University of Jena.}

\bibliographystyle{JHEP}
\bibliography{adjointmodes.bib}

\typeout{get arXiv to do 4 passes: Label(s) may have changed. Rerun}

\end{document}